\documentclass[runningheads,a4paper]{llncs}

\usepackage{amssymb}
\usepackage{graphicx}

\usepackage{url}
\urldef{\mailsa}\path|huangzh2.13@sem.tsinghua.edu.cn|
\urldef{\mailsb}\path|{aoxiang, liuyang17z, zuol, heqing}@ict.ac.cn|
\newcommand{\keywords}[1]{\par\addvspace\baselineskip
\noindent\keywordname\enspace\ignorespaces#1}

\usepackage{times}
\usepackage{amsmath,epsfig}
\usepackage{amssymb}
\usepackage{epstopdf}
\usepackage{graphicx,subfigure}
\usepackage{enumerate}
\usepackage{amsmath}
\usepackage{amscd}
\usepackage{rotating}
\usepackage{multirow}
\usepackage{url}
\usepackage{extarrows}
\usepackage{booktabs}
\usepackage{balance}
\usepackage{color}
\usepackage{colortbl}
\usepackage{caption}
\usepackage{amsmath}
\usepackage{float}
\usepackage{bm}

\usepackage[linesnumbered,vlined,ruled]{algorithm2e}

\newcommand{\bluefont}[1]{{\color{blue}{#1}}}

\newcommand{\ev}[1]{{\small \textsf{#1}}}

\begin{document}

\mainmatter  

\title{Free-rider Episode Screening via Dual Partition Model}

\titlerunning{Free-rider Episode Screening via Dual Partition Model}

%
%
\author{Xiang Ao$^{1, 2}$
\and Yang Liu$^{1, 2}$
\and Zhen Huang$^{3}$\thanks{This work was done when Zhen was visiting Institute of Computing Technology, CAS.} \and Luo Zuo$^{1, 2}$ \and Qing He$^{1, 2}$}
\authorrunning{Lecture Notes in Computer Science: Authors' Instructions}
%
\institute{$^{1}$Key Lab of Intelligent Information Processing of Chinese Academy of Sciences (CAS), \\ Institute of Computing Technology, CAS, Beijing 100190, China.\\
$^{2}$University of Chinese Academy of Sciences, Beijing 100049, China.\\
$^{3}$Tsinghua University, Beijing, China.\\
\mailsb\\
\mailsa\\
}

\authorrunning{X. Ao et al.}
%
%
\toctitle{Free-rider Episode Screening via Dual Partition Model}
\tocauthor{Free-rider Episode Screening via Dual Partition Model}

\maketitle

\begin{abstract}
One of the drawbacks of frequent episode mining is that overwhelmingly many of the discovered patterns are redundant. Free-rider episode, as a typical example, consists of a real pattern doped with some additional noise events. Because of the possible high support of the inside noise events, such free-rider episodes may have abnormally high support that they cannot be filtered by frequency based framework.
An effective technique for filtering free-rider episodes is using a partition model to divide an episode into two consecutive subepisodes and comparing the observed support of such episode with its expected support under the assumption that these two subepisodes occur independently.
In this paper, we take more complex subepisodes into consideration and develop a novel partition model named EDP for free-rider episode filtering from a given set of episodes. It combines (1) a dual partition strategy which divides an episode to an underlying real pattern and potential noises;
(2) a novel definition of the expected support of a free-rider episode based on the proposed partition strategy.
We can deem the episode interesting if the observed support is substantially higher than the expected support estimated by our model.
The experiments on synthetic and real-world datasets demonstrate EDP can effectively filter free-rider episodes compared with existing state-of-the-arts.

\keywords{Episode dual partition  $\cdot$ Interesting pattern discovery $\cdot$ Episode mining $\cdot$ Sequence mining}
\end{abstract}


\section{Introduction}\label{sec:intro}

One of the defects in frequent pattern mining is that there are abundant redundant patterns in the very large number of output patterns~\cite{vreeken2014interesting}. As a result, how to effectively reduce redundancy of the output becomes an essential problem of current research~\cite{webb2010self,mampaey2012summarizing,lam2014mining,tatti2015ranking,petitjean2016skopus,ibrahim2016discovering,bertens2015keeping,fowkes2016subsequence}. Frequent episode mining~\cite{fem:mannila1997discovery}~(FEM for short), as one of the sub-topics of frequent pattern mining, which aims at discovering frequently appeared ordered sets of events from a single symbol~(event) sequence, is facing the similar problem as well.
Moreover, FEM algorithms usually suffer from inefficient processing since checking whether a sequence covers a general episode is an \textbf{NP}-hard problem~\cite{fem:tatti2011mining}.
It makes redundancy reduction in FEM particularly crucial because users will be reluctant to obtain redundant and useless patterns after a long time waiting.

In this paper, our purpose is to reduce redundancy of frequent episodes. Specifically, we focus on screening \emph{free-rider episodes} and identifying underlying patterns from a given set of frequent episodes.
Here free-rider episode, as a prototypical example of redundancy in episodes, consists of a real pattern doped with additional noise events.

While filtering patterns to reduce redundancy is well-studied for itemsets, it is still underexplored for episodes.
The most straightforward approach to solve such problem is to compare observations against expectations predicted by independence model~\cite{gwadera2005reliable,tatti2014discovering,laxman2005discovering}.
Interestingly enough, episodes occurrences in a single sequence may be dependent to each other, and hence support is no longer a sum of independent variables~\cite{tatti2015ranking}.
A more effective manner is to utilize partition models in which an episode is divided into subepisodes and the observed support is compared with the expected support produced by the model under specialized assumptions~\cite{tatti2015ranking,petitjean2016skopus}. However, constructing a good partition model has significant challenges that existing approaches present some limitations, and we summarize them as follows.

\begin{itemize}
\item Noise events could be doped \emph{before}, \emph{after}, or \emph{inside} a real pattern. For example, \ev{a}$\rightarrow$$x$$\rightarrow$\ev{b} could be a free-rider episode, where \ev{a}$\rightarrow$\ev{b} is the real pattern and $x$ is the noise. Here \ev{a}$\rightarrow$\ev{b} is a non-prefix subepisode of \ev{a}$\rightarrow$$x$$\rightarrow$\ev{b}. However,  most existing partition models fail to take non-prefix subepisodes into account. For example, in~\cite{tatti2015ranking},  \ev{a}$\rightarrow$$x$$\rightarrow$\ev{b} can only be divided into two consecutive subepisodes, e.g. $\{$\ev{a}$\rightarrow$$x$, \ev{b}$\}$ or $\{$\ev{a}, $x$$\rightarrow$\ev{b}$\}$.
\item It is challenging to define the expected support of an episode in one long sequence.
The major reason is that the support of an episode is defined as its number of occurrences in a sequence, and the events in such sequence may be dependent to each other.
Hence we have to carefully use independence assumptions in this scenario.
However, independence assumptions are widely used in redundancy reduction in database of itemsets or sequences~\cite{webb2010self,tatti2015ranking}.
In that problem, the support of pattern is usually defined as the number of the transactions/sequences covering the pattern, and each individual transaction/sequence is naturally regarded independent with each other. As a result, it is difficult to adapt traditional ideas to our problem.

\item The search space of partitions suffers from ``combination explosion'' as it increases exponentially to the number of events in an episode.  Hence pruning strategies are needed to perform an efficient filtering.
\end{itemize}

To address the challenges and overcome the limitations of the existing methods, we propose a novel partition model named EDP~(\textbf{E}pisode \textbf{D}ual \textbf{P}artition).

In EDP, we partition the events of an episode into two distinct categories, called \emph{informative} and \emph{random} respectively, and we make the assumption that informative events form the real pattern while random events are noises. Here we do not restrict the position of either informative or random events and thus non-prefix subepisodes are taken into consideration.
Then we build a \emph{generative model} for random sequences which are considered producing free-rider episodes by the following ways. First, we fix the occurrences of informative events according to their positions in the original sequence. Second, we generate random events with an assumption that they are independent of potential real patterns.
Next, based on these random sequences, we define the expected support of a free-rider episode and adopt an automaton based algorithm to compute.
Finally we deem an episode redundant if its observed support can be explained by the estimated expected support. Otherwise, it will be an informative pattern. The experiments on both synthetic and real-world dataset about finance and text demonstrate the effectiveness of our model.

\section{Related Work}\label{sec:survey}\vspace{-2mm}
We categorize related papers in the literature as follows.

\textbf{Mining frequent episodes.} Mining frequent episodes from event sequence was first introduced by Mannila et al.~\cite{fem:mannila1997discovery} where they defined episodes as directed acyclic graphs and considered two kinds of methods counting support, i.e. sliding window and minimal occurrence.
Mining general episodes can be intricate and computing-intensive because discovering whether a sequence covers a general episode is \textbf{NP}-hard~\cite{fem:tatti2011mining}. As a result, efforts have been focused on mining subclasses of episodes, such as serial episodes~\cite{ao2014discovering,fem:ao2015online,ao2017mining}, closed episodes~\cite{fem:tatti2011mining,fem:tatti2010mining}, episodes with unique labels~\cite{achar2012discovering,pei2006discovering}.

\textbf{Filtering episodes based on expectation.} Our EDP model belongs to such category. That is to define the expected support for an episode and compare with its observed support for filtering or ranking tasks. Some existing efforts compare the observed support against the independence model~\cite{gwadera2005reliable,tatti2014discovering,laxman2005discovering}. In such models, they use a null hypothesis that the data are generated by a stationary and independent stochastic process.
Recent work estimated the expected support by considering different partitions of episodes.
Tatti~\cite{tatti2015ranking} divided an episode into two consecutive subepisodes and downplayed the importance of such episode if its observed support could be explained by the two subepisodes. Such model failed to take non-prefix subepisodes into account when perform partitioning. Therefore it may be difficult to detect free-rider episodes when noise events are interweaved with a real pattern.
Some idea for filtering sequential patterns can be adapted to episodes as well. For example, Petitjean et al. proposed SkOPUS~\cite{petitjean2016skopus} in which re-ordering candidates were made from two subpattern partitions, and the expected support was defined based on the mean of support of all possible candidates in the original sequences.
However, such work may distort the expected support as some re-orderings may be surprisingly rare in the original sequence.

\textbf{Episode set mining.} There is also another kind of efforts to score a set of episodes which best explain the data. Most of them are Minimum Description Length (MDL) based approaches for scoring a set
of patterns relative to a dataset as well as a heuristic search method to construct the set~\cite{tatti2012long,lam2014mining,ibrahim2016discovering,fowkes2016subsequence,bhattacharyya2017efficiently}. Our work is different with such episode set mining
since episode set mining methods select episodes based on how well we can describe the data using the episodes while our purpose is to filter free-rider episodes from real patterns based on how well their support can be explained by random sequences.

\section{Definitions and Problem Statement}\label{sec:definition}\vspace{-2mm}
In this section, we introduce preliminaries, definitions and problem statement.
Let $\Omega$ $=$ $\{e_1, e_2, \dots, e_m\}$ be a finite event alphabet. We will use $\Omega$ $=$ $\{\ev{a},\ev{b},\ev{c},\ev{d}\}$ for all examples in this paper. An event sequence $\bm{S}$ $=$ $\langle$$(E_1, t_1)$, $(E_2, t_2)$, $\cdots$, $(E_n, t_n)$$\rangle$, is an ordered sequence of events,
where each $E_{i} \subseteq \Omega$ consists of all events associated with time stamp $t_i$, $1 \leqslant i \leqslant n$, and $t_j<t_k$ for any $1\leqslant j < k \leqslant n$.
In addition, $n$ is the \emph{length} of event sequence $\bm{S}$, denoted by $\mathrm{len}(\bm{S})=n$.

\begin{figure} [!htbp]
\centering\vspace{-5mm}
\includegraphics[scale=0.4]{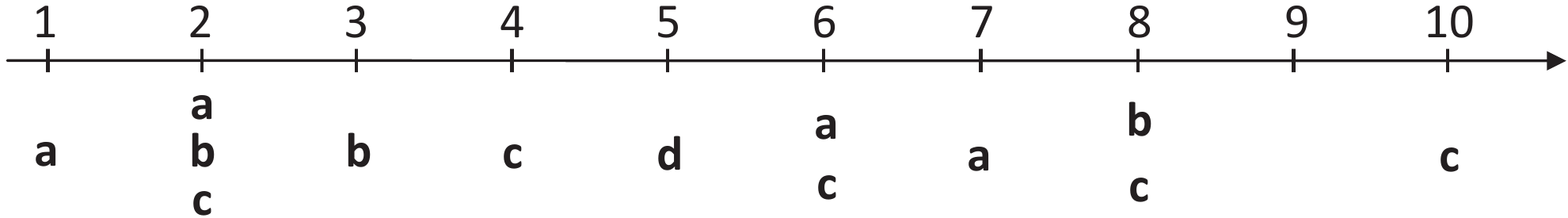}
\vspace{-3mm}
\caption{The example event sequence.}
\label{fig:seq}
\end{figure}

For example, Fig.~\ref{fig:seq} shows an event sequence $\bm{S}$ with $\mathrm{len}(\bm{S})$ $=$ $10$. $\bm{S}$ will be used as the running example in this paper.

\vspace{-5mm}

An episode~(refer to serial episode in this paper) $\alpha$ is a totally ordered events in the form of $e_{\alpha_1}$$\rightarrow$$e_{\alpha_2}$$\rightarrow$$\cdots$$\rightarrow$$e_{\alpha_k}$ where $\alpha_i$ $\in$ $[1, m]$ and thus $e_{\alpha_i}$ $\in$ $\Omega$ for all $1 \leqslant i \leqslant k$. We can abbreviate $\alpha$ $=$ $e_{\alpha_1}$$\rightarrow$$e_{\alpha_2}$$\rightarrow$$\cdots$$\rightarrow$$e_{\alpha_k}$ as $e_{\alpha_1}$$e_{\alpha_2}$$\cdots$$e_{\alpha_k}$. We will use such abbreviation to represent episode in this paper unless otherwise specified.
The \emph{length} of an episode is defined as the number of events in the episode.
An episode of length $k$ is called a $k$-episode. We call an episode with length $0$ as an \emph{empty episode}, and denote it by $\emptyset$.
A minimal occurrence window of $\alpha$ is a time-window $[t_s, t_e]$ which contains an occurrence of $\alpha$, such that no proper sub-window of it contains another occurrence of $\alpha$. $t_s$ and $t_e$ are called \emph{start time} and \emph{end time}, respectively. Usually there is an additional threshold of maximal window size $\delta$ such that $t_e$ $-$ $t_s$ $<$ $\delta$.
The set of all distinct minimal occurrence window of an episode $\alpha$ is denoted by $\mathrm{moSet}(\alpha)$.

For example, $\mathrm{moSet}(\ev{abc})$ $=$ $\{[2, 4], [7, 10]\}$ if $\delta$ $=$ $4$ in Fig.~\ref{fig:seq}. Though $[1, 4]$ contains an occurrence of episode \ev{abc}, it is not a minimal occurrence window of \ev{abc} because it subsumes the time window $[2, 4]$ which contains another occurrence of \ev{abc}.

The support of an episode $\alpha$, denoted by $\mathrm{sp}(\alpha)$, is defined as the number of distinct minimal occurrence windows of $\alpha$ in sequence $\bm{S}$, i.e. $\mathrm{sp}(\alpha)$ $=$ $|\mathrm{moSet}(\alpha)|$.
We call an episode $\alpha$ \emph{frequent} if its support passes a user-specific \emph{minimum support threshold} $\mathit{min}\_\mathit{sup}$ $>$ $0$, i.e. $\mathrm{sp}(\alpha)$ $\geqslant$ $\mathit{min}\_\mathit{sup}$. Otherwise, it is infrequent.
Consider two episodes $\alpha$ $=$ $e_{\alpha_1}$ $\cdots$ $e_{\alpha_k}$ and $\beta$ $=$ $e'_{\alpha_{1}}$ $\cdots$ $e'_{\alpha_{m}}$ where $m$ $\leqslant$ $k$, the episode $\beta$ is a \emph{subepisode} of $\alpha$, denoted by $\beta$ $\preceq$ $\alpha$, if and only if there exist $m$ integers $1\leqslant{i_1}\leqslant{i_2}\cdots\leqslant{i_m}\leqslant{k}$ such that $e'_{\alpha_{j}}$ $=$ $e_{\alpha_{i_j}}$ where $1$ $\leqslant$ $j$ $\leqslant$ $m$. We call an episode $H_j^{\alpha} = e_{\alpha_{1}} \cdots e_{\alpha_{j}}$ a \emph{$j$-prefix} episode of $\alpha$ if $0 \leqslant j \leqslant k$. Specifically, $0$-prefix episode of any non-empty episode $\alpha$ is the empty episode $\emptyset$.

For example, all possible subepisodes of $\alpha = \ev{abc}$ include $\emptyset$, \ev{a}, \ev{b}, \ev{c}, \ev{ab}, \ev{ac}, \ev{bc} and \ev{abc}.
All prefix episodes of $\alpha$ $=$ $\ev{abc}$ are $H_0^{\ev{abc}}$ $=$ $\emptyset$, $H_1^{\ev{abc}}$ $=$ $\ev{a}$, $H_2^{\ev{abc}}$ $=$ $\ev{ab}$, $H_3^{\ev{abc}}$ $=$ $\ev{abc}$.

Next, we define the concept of free-rider episode and give a formal statement of the problem.
We assume the knowledge of generative rules behind real patterns in advance and give the following assumption: A \emph{generative rule} $\alpha \Rightarrow e$ states that an event $e$ will be more likely to occur after an occurrences of an episode $\alpha$.

For example, $\ev{ab} \Rightarrow \ev{c}$ is the generative rule indicates that the probability of occurrence of event $\ev{c}$ will increase in some time stamps after episode $\ev{ab}$ occurs.
With the defined generative rule, free-rider events in an episode can be seen as those events that no generative rule is associated with.

\textbf{Free-rider Event.} A free-rider event $e_{\alpha_i}$ in an episode $\alpha$ $=$ $e_{\alpha_1}$$\cdots$$e_{\alpha_k}$, where $1$ $\leqslant$ $i$ $\leqslant$ $k$, is an event that there is no generative rule associated with in the form as follows.
\begin{enumerate} \label{rules}
  \item $\beta$$\Rightarrow$$e_{\alpha_i}$, where $\beta$ is a non-empty subepisode of $(i-1)$-prefix episode of $\alpha$, i.e., $\beta$ $\preceq$ $H_{i-1}^{\alpha}$ or
  \item $\beta'$$\Rightarrow$$e_{\alpha_j}$, where $j>i$ and $\beta'$ containing $e_{\alpha_i}$ is a non-empty subepisode of $(j-1)$-prefix episode of $\alpha$, i.e., $e_{\alpha_i}$ $\in$ $\beta'$$\preceq$$H_{j-1}^{\alpha}$.
\end{enumerate}

For instance, consider episode $\alpha = \ev{abc}$, $\ev{b}$ will not be a free-rider event if $\ev{a} \Rightarrow \ev{b}$ holds or if $\ev{ab} \Rightarrow \ev{c}$ holds. However, generative rule $\ev{ac} \Rightarrow \ev{b}$ has no impact on whether $\ev{b}$ is a free-rider event of the episode \ev{abc} or not. 
In other words, the event order in $\alpha$ should be respected.
Then we can describe free-rider episodes as follows.

\textbf{Free-rider Episode.} An episode $\alpha$ $=$ $e_{\alpha_1}$$\cdots$$e_{\alpha_k}$ is a \emph{free-rider episode} if it contains at least one free-rider event $e_{\alpha_i}$ where $1\leqslant{i}\leqslant{k}$.

We argue that filtering free-rider episodes by definition is intricate since it may be hard to identify free-rider events by checking these implicit generative rules. To close the gap, however, we devise a novel expected support of a free-rider episode and take \emph{Lift}, which is the ratio of observed support to the expectation, as the measure to determine whether an episode is a free-rider episode or not. We can then filter the episodes whose Lift values fail to exceed the minimum lift threshold $\mathit{min}\_\mathit{lift}$ and rank the rest by Lift with a descending order as their interestingness scores.
Hence, the problem of free-rider episode screening in this paper is formulated as follows.

\textbf{Problem Statement of Free-rider Episode Screening.} Given an event sequence $\bm{S}$ and a set of frequent episodes on $\bm{S}$, the free-rider episode screening problem is to rank the episodes with the Lift produced by the EDP model and filter the episodes whose Lift values are less than $\mathit{min}\_\mathit{lift}$.

\section{The EDP Model} \label{sec:model}
The key ingredient to our problem is to compute the expected support of a free-rider episode, and we thus propose the EDP model.
It involves an episode partition strategy, a generative model for random sequences based on episode partitions, and an expected support definition of a free-rider episode based on the generated random sequences. The inherent idea behind our model is that if the support of an episode can be simulated by that from a group of random sequences, it cannot be a real pattern.

\subsection{Episode Partition Strategy}

We first detail the episode partition strategy in EDP. Since pattern and noise events in an episode may interweave in complex ways we first make the following assumption to simplify the problem. Specifically, we assume that an event of an episode will not simultaneously be a part of real pattern and a noise event. Such hard assignment could simplify the processes of sequence generation and expectation calculation that will be introduced in the following parts. Formally, given an episode $\alpha$$=$$e_{\alpha_1} \dots e_{\alpha_k}$, there is no $i$, $j$ $\in$ $[1, k]$ and $i$ $\neq$ $j$ such that $e_{\alpha_i}$$=$$e_{\alpha_j}$ while $e_{\alpha_i}$ is a free-rider event and $e_{\alpha_j}$ is not.

Take the episode $\ev{abb}$ as an example. There are two event \ev{b} in such episode. We assume it is not possible that the first \ev{b} is a noise event while the second \ev{b} is not.
The practical usefulness of our model will not be limited by this assumption, as shown in our experiments.
Under such assumption, we have the following partition strategy.

\begin{definition}\label{def:ialpha}
\textbf{\emph{(Dual Partition)}} Let $\Omega_{\alpha}$ be the set of event alphabet of an episode $\alpha$. We divide $\Omega_{\alpha}$  into two distinct parts called \emph{informative} and \emph{random} events, which are denoted by $\mathcal{I}_{\alpha}$ and $\overline{\mathcal{I}_{\alpha}}$, respectively. We assume the events in $\mathcal{I}_{\alpha}$ form the real pattern, and the events in $\overline{\mathcal{I}_{\alpha}}$ are regarded as noises. Then the episode $\alpha$ is partitioned as a subepisode of $\alpha$ which consists of events in $\mathcal{I}_{\alpha}$, and noise events in $\overline{\mathcal{I}_{\alpha}}$.
\end{definition}

Such partition indeed takes non-prefix subepisodes of $\alpha$ into account when performing partitioning. For example, consider an episode $\alpha$ $=$ \ev{abcd}, we have $\Omega_{\alpha}$ $=$ $\{$\ev{a}, \ev{b}, \ev{c}, \ev{d}$\}$. $\Omega_{\alpha}$ can be partitioned as $\mathcal{I}_{\alpha}$ $=$ $\{\ev{a}, \ev{c}\}$ and $\overline{\mathcal{I}_{\alpha}}$ $=$ $\{\ev{b}, \ev{d}\}$ or $\mathcal{I}_{\alpha}$ $=$ $\{\ev{a}, \ev{b}, \ev{d}\}$ and $\overline{\mathcal{I}_{\alpha}}$ $=$ $\{\ev{c}\}$. Under these two partitions, $\alpha$ can be divided into a subepisode \ev{a}\ev{c} and a noise event set $\{\ev{b}, \ev{d}\}$ or \ev{a}\ev{b}\ev{d} and $\{\ev{c}\}$. Both \ev{a}\ev{c} and \ev{a}\ev{b}\ev{d} are non-prefix subepisodes of \ev{abcd}.



%




\subsection{Generate Random Sequences} \label{fixing}
Next we devise a generative model~(denoted by $\mathcal{M}_{\mathcal{I}_{\alpha}}$), for a given episode $\alpha$, generating random sequences based on the aforementioned episode dual partition $\mathcal{I}_{\alpha}$ and $\overline{\mathcal{I}_{\alpha}}$.

The basic idea for $\mathcal{M}_{\mathcal{I}_{\alpha}}$ is that we consider the potential real pattern, which consists of informative events, follows implicit generative rules. 
Their occurrences in the original event sequence are thus meaningful. While the random events, which are independent of the real patterns, could occur at any time stamp.
Hence the model generates random sequences as follows. First, the length of every random sequence is the same as the original event sequence. Second, the model produces informative events according to their occurrences in the original sequence. Third, the random events are generated independently over all possible time stamps.

Formally, we first define the probability of generating an event $e$ at a specific time stamps $t_j$ under $\mathcal{M}_{\mathcal{I}_{\alpha}}$ as follows.

\begin{equation}\scriptsize \label{equation-p}
P(e|\mathcal{M}_{\mathcal{I}_{\alpha}}, t_j) = \begin{cases}
      1 & {e \in \mathcal{I}_{\alpha}} ~ \& ~   { e \in E_j } \\
      0 & {e \in \mathcal{I}_{\alpha}} ~ \& ~  { e \not\in E_j } \\
      p_{\mathit{ind}}(e) & {e \in \overline{\mathcal{I}_{\alpha}}}
   \end{cases} \\
\end{equation}


In Formula~\ref{equation-p}, $E_j$ is the event set associated with time stamp $t_{j}$ in the original event sequence $\bm{S}$, and $p_{\mathit{ind}}(e)$ is the occurring probability\footnote{$p_{\mathit{ind}}(e)$ can be calculated by $\frac{\mathrm{sp}(e)}{\mathrm{len}(\bm{S})}$, where $\mathrm{sp}(e)$ is the support of event $e$ in $\bm{S}$.} of event $e$ in $\bm{S}$ .

Next we define the probability for generating an event set $E'_j$ based on $\mathcal{M}_{\mathcal{I}_{\alpha}}$ at a specific time stamp $t_j$ as follows.
\begin{equation} \scriptsize
\label{equation-pE}
P(E'_j|\mathcal{M}_{\mathcal{I}_{\alpha}}, t_j) = \prod_{e \in E'_j} P(e|\mathcal{M}_{\mathcal{I}_{\alpha}}, t_j)\prod_{e \in \Omega_{\alpha}\setminus E'_j}(1 - P(e|\mathcal{M}_{\mathcal{I}_{\alpha}}, t_j))
\end{equation}



We can generate event set for every time stamp with Formula~\ref{equation-pE} and eventually obtain a random sequence by assuming every event set is independent to each other. Here we can safely use the independent assumption here since we attempt to generate random sequences without specialized rules.
Remember that the length of a random sequence is the same as the length of $\bm{S}$. Therefore, the probability to generate a specific random sequence $\hat{\bm{S}}$ can be given as Eq.~\ref{eq:generateS}, where $n$ denotes the length of $\bm{S}$. 
\vspace{-2mm}
\begin{equation} \scriptsize\label{eq:generateS}
P(\hat{\bm{S}}|\mathcal{M}_{\mathcal{I}_{\alpha}}, n) = \prod_{j =1}^{n} P(E'_j|\mathcal{M}_{\mathcal{I}_{\alpha}}, t_j)
\end{equation}

\begin{figure} [!htbp]
\centering\vspace{-3mm}
\includegraphics[scale=0.4]{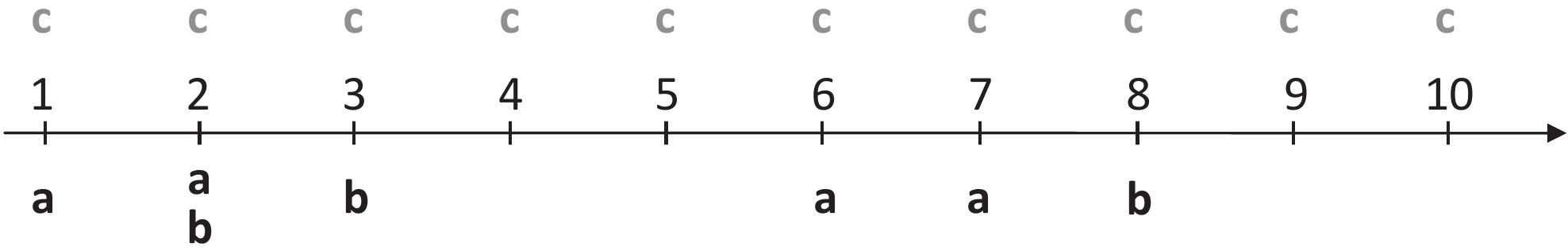}
\vspace{-1mm}
\caption{The generative model $\mathcal{M}_{\{\ev{a}, \ev{b}\}}$ for episode \ev{abc}. Occurrences of event \ev{a} and \ev{b} are fixed according to their occurrences in the original sequence. Event \ev{c} is generated with its occurring probability in the original sequence.}
\label{fig:generative}\vspace{-10mm}
\end{figure}

For example, consider the event sequence shown in Fig.~\ref{fig:seq} and the episode $\alpha = \ev{a}\ev{b}\ev{c}$. Suppose $\mathcal{I}_{\alpha} = \{\ev{a}, \ev{b}\}$ and $\overline{\mathcal{I}_{\alpha}} = \{\ev{c}\}$. The generative model $\mathcal{M}_{\{\ev{a}, \ev{b}\}}$ fixes all the occurrences of event \ev{a} and \ev{b} as in the sequence shown in Fig.~\ref{fig:seq} and generates event \ev{c} by $p_{\mathit{ind}}(\ev{c})$ $=$ $0.5$~($\mathrm{sp}(\ev{c})=5$ and $\mathrm{len}(\bm{S})=10$). The illustration of $\mathcal{M}_{\{\ev{a}, \ev{b}\}}$ is shown as Fig.~\ref{fig:generative} in which the fixed informative events are shown under the time stamps while the random event \ev{c} is shown above with the grey font.
For probability of the candidate event set at each time stamp, it is the multiplication of all possible events in it, e.g. $P(E'_7 = \{\ev{a}, \ev{c}\}|\mathcal{M}_{\{\ev{a}, \ev{b}\}}, 7)$ $=$ $P(E'_7 = \{\ev{a}\}|\mathcal{M}_{\{\ev{a}, \ev{b}\}}, 7)$ $=$ $0.5$ and $P(E'_9 = \{\ev{c}\}|\mathcal{M}_{\{\ev{a}, \ev{b}\}}, 9)$ $=$ $P(E'_9 = \{\emptyset\}|\mathcal{M}_{\{\ev{a}, \ev{b}\}}, 9)$ $=$ $0.5$.
Though $E_{9}$ has no event in the original sequence, the model may generate events based on the dual partition.


\subsection{Expected Support Definition}

Next we define the expected support of a free-rider episode on these random sequences.
Recall that given an episode $\alpha$, EDP generates random sequences with events in $\Omega_{\alpha}$ based on the model $\mathcal{M}_{\mathcal{I}_{\alpha}}$, and we assume, in $\mathcal{M}_{\mathcal{I}_{\alpha}}$, events belonging to $\mathcal{I}_{\alpha}$ have implicit generative rules but events in $\overline{\mathcal{I}_{\alpha}}$ do not.
Hence based on our definition every event $e'$ in $\overline{\mathcal{I}_{\alpha}}$ will be a free-rider event to $\alpha$ in each generated sequence and $\alpha$ in these sequences will become a free-rider since $e' \in \alpha$. As a consequence, denoted the set of random sequences generated by a given generative model $\mathcal{M}_{\mathcal{I}_{\alpha}}$ as $\mathrm{Seq}(\mathcal{M}_{\mathcal{I}_{\alpha}})$, we match the minimal occurrences of $\alpha$ on each generated sequence $\hat{\bm{S}} \in \mathrm{Seq}(\mathcal{M}_{\mathcal{I}_{\alpha}})$ and define the expected support of $\alpha$, which is regarded as a free-rider episode, as follows.

\vspace{-1mm}
\begin{equation} \scriptsize
\label{eq:expbymodel}
\mathbb{E}_{\hat{\bm{S}}\sim{P(\hat{\bm{S}} | \mathcal{M}_{\mathcal{I}_{\alpha}}, n)}}[\mathrm{sp}(\alpha | \hat{\bm{S}})] = \sum_{\hat{\bm{S}} \in \mathrm{Seq}(\mathcal{M}_{\mathcal{I}_{\alpha}})} \mathrm{sp}(\alpha | \hat{\bm{S}}) \cdot P(\hat{\bm{S}}|\mathcal{M}_{\mathcal{I}_{\alpha}}, n)
\end{equation}

where $\mathrm{sp}(\alpha | \hat{\bm{S}})$ denotes the support of the episode $\alpha$ in a generated sequence $\hat{\bm{S}}$.

It is no doubt different generative models will derive different expected support for an episode $\alpha$. To minimize the false positive, we define the formal expected support as the maximum expected support value produced by a specific generative model.
Hence we define the expected support of an episode $\alpha$, denoted as $\mathrm{ExpSup}(\alpha)$, as follows.

\begin{equation}\scriptsize \label{eq:expectedsupport}
\mathrm{ExpSup}(\alpha) = \max_{\mathcal{M}_{\mathcal{I}_{\alpha}}} \mathbb{E}_{\hat{\bm{S}}\sim{P(\hat{\bm{S}} | \mathcal{M}_{\mathcal{I}_{\alpha}}, n)}}[\mathrm{sp}(\alpha | \hat{\bm{S}})]
\end{equation}

With such definition, we adopt the following $\mathit{Lift}(\alpha)$ to measure the deviation between the observed support of an episode $\alpha$ and its expectation and regard it as a free-rider episode if $\mathit{Lift}(\alpha)$ does not exceed the minimum lift threshold $\mathit{min}\_\mathit{lift}$.
\vspace{-1mm}
\begin{equation} \scriptsize\label{equation-lift}
\mathit{Lift}(\alpha) = \frac{\mathrm{sp}(\alpha)}{\mathrm{ExpSup}(\alpha)}
\end{equation}

Under this formula, we actually make the following assumption: For an episode $\alpha$ in an event sequence $\bm{S}$, if we could find a generative model producing free-rider episodes in the same form of $\alpha$ by EDP which can generate $\alpha$ more frequently or close to the observed support of $\alpha$ in $\bm{S}$, then $\alpha$ in $\bm{S}$ will be redundant.

\subsection{Time Complexity of EDP}

In this subsection, we analyze the time complexity of the proposed EDP.
Though the search space of possible partitions of an episode $\alpha$ is  $2^{|\Omega_{\alpha}|}$, where $\Omega_{\alpha}$ is the alphabet of $\alpha$, we could stop calculation and screen $\alpha$ if we find one $\mathcal{M}_{\mathcal{I}_{\alpha}}$ such that the value of $\frac{\mathrm{sp}(\alpha)}{\mathbb{E}_{\hat{\bm{S}}\sim{P(\hat{\bm{S}} | \mathcal{M}_{\mathcal{I}_{\alpha}}, n)}}[\mathrm{sp}(\alpha | \hat{\bm{S}})]}$ is clearly less than $\mathit{min}\_\mathit{lift}$.
It indicates there exists one group of random sequences that provides free-rider episodes of $\alpha$ can explain $\alpha$'s observed support in the input sequence. As a result, we can explore pruning techniques based on such property.

For a real pattern, on the other hand, we need to enumerate all partitions to ascertain its interestingness. Hence the time complexity will be $O(2^{|\Omega_{\alpha}|})$. However, in our EDP each partition has no impact on others, the process can thus be highly parallelized without hurting the performance. The experimental results will show EDP can achieve significant speedup in running time if we distribute the checking into multi-processes.

\section{Expected Support Calculation} \label{expected}
So far our problem is how to compute the expected support of an episode $\alpha$ given a generative model $\mathcal{M}_{\mathcal{I}_{\alpha}}$.
In this section, we devise an algorithm computing the expected support defined in Eq.~\ref{eq:expbymodel} with the help of automaton.

\subsection{Tracking Minimal Occurrences with Automaton}
Automaton is an effective tool to track minimal occurrence of episode in sequences. Given an episode $\alpha$ $=$ $e_{\alpha_1}$$e_{\alpha_2}$$\cdots$$e_{\alpha_k}$, we can obtain a unique automaton $\mathcal{A}_{\alpha}$ in which every node~(also known as state) denotes a prefix subepisode of $\alpha$, namely $H_{i}^{\alpha}$ where $0$ $\leqslant$ $i$ $\leqslant$ $k$. The label on each edge denotes the event that can trigger state transition. We denote the transition function by $T(H, E)$ where $H$ is the current state of an automaton and $E$ is an event set. The first state and the last state of an automaton are called the source state and the sink state, respectively.
Since minimal occurrences may overlap to each other on the event sequence, we need to simultaneously manage multiple automatons which do not reach the sink state during scanning the sequence. We call them active automatons.

\subsection{The Algorithm}
The pseudo code of expected support calculation is shown as Algorithm~\ref{alg:expect}.

In our algorithm, we do not need to generate any specific random sequence. Instead, we adopt the probability of event set generation and the probability of the appearance of active automatons to estimate the expected support defined in Eq.~\ref{eq:expbymodel}.
Since $\mathcal{M}_{\mathcal{I}_{\alpha}}$ may generate multiple different event sets for each time stamp, we need to simultaneously manage more than one possible active automaton lists and consider all possible event sets for every time stamp.
In more detail, we use a variable $L_{i}$ to store active automatons at time stamp $t_{i}$ and first initialize $L_{0}$ $=$ $\{H_{0}^{\alpha}\}$.
Since $L_{0}$ is determined, its appearance probability is one, and the list of active automatons, denoted as $\mathcal{L}_{0}$, contains the only element $L_0$~(Line~\ref{alg-l2}--\ref{alg-l3}).

\vspace{-3mm}
\begin{algorithm}[!htpb]\scriptsize
\caption{Expected support calculation given $\mathcal{M}_{\mathcal{I}_\alpha}$}
\label{alg:expect}
\KwIn{
$\alpha$: a $k$-episode

$\mathcal{M}_{\mathcal{I}_\alpha}$: a generative model constructed by episode partition $\mathcal{I}_{\alpha}$

$\bm{S}$: the original event sequence
}
\KwOut{$\mathbb{E}_{\hat{\bm{S}}\sim{P(\hat{\bm{S}} | \mathcal{M}_{\mathcal{I}_{\alpha}}, n)}}(\mathrm{sp}(\alpha | \hat{\bm{S}}))$: the expected support of an episode $\alpha$ given a generative model $\mathcal{M}_{\mathcal{I}_\alpha}$}

    initialize $L_{0}$ $=$ $\{H_0^{\alpha}\}$, $\mathcal{L}_{0}$ $=$ $\emptyset$,  $P(L_0) = 1.0$, $\mathbb{E}_{\hat{\bm{S}}\sim{P(\hat{\bm{S}} | \mathcal{M}_{\mathcal{I}_{\alpha}}, n)}}(\mathrm{sp}(\alpha | \hat{\bm{S}}))$ $=$ $0$ \label{alg-l2}

    $\mathcal{L}_{0}$ $\gets$ $\mathcal{L}_{0}$ $\cup$ $\{L_{0}\}$ \label{alg-l3}

    \For { $i$ $=$ $1$ to $\mathrm{len}(\bm{S})$} { \label{alg-l4}

        $\mathcal{L}_{i}$ $=$ $\emptyset$ \label{alg-l5}

       \ForEach{possible active automaton list $L_{i-1}$ $\in$ $\mathcal{L}_{i-1}$}{\label{alg-l6}
            $P(L_{i})$ $=$ $0$

            \ForEach{possible event set $E_i'$ produced by $\mathcal{M}_{\mathcal{I}_{\alpha}}$}{\label{alg-l7} 

                \ForEach{active $\mathcal{A}(\alpha)$ $\in$ $L_{{i-1}}$}{\label{alg-l8}

                    $H_{c}$ $\gets$ current state of $\mathcal{A}(\alpha)$ \label{alg-l9}

                    $H_t$ = $T(H_c, E_{i}')$ \label{alg-l10}

                    \If{TRANSIT i.e. $H_t \neq H_c$}{\label{alg-l11}
                        \If{SINK i.e. $H_t = H_k^{\alpha}$}{\label{alg-l12}
                            increase $\mathbb{E}_{\hat{\bm{S}}\sim{P(\hat{\bm{S}} | \mathcal{M}_{\mathcal{I}_{\alpha}}, n)}}(\mathrm{sp}(\alpha | \hat{\bm{S}}))$ by $P(L_{i-1})$ $\cdot$  $P(E'_i | \mathcal{M}_{\mathcal{I}_{\alpha}}, t_i)$ \label{alg-l13}
                        }\Else{
                            add a copy of $\mathcal{A}(\alpha)$ whose current state is $H_{t}$ to $L_{i}$ \label{alg-l15}
                        }
                        \If{SOURCE i.e. $H_c = H_0^{\alpha}$}{\label{alg-l16}
                            add an initialized $\mathcal{A}(\alpha)$ whose current state is $H_{0}^{\alpha}$ to $L_{i}$ \label{alg-l17}
                        }
                    }
                }

     \ForAll{$\mathcal{A}(\alpha)$ $\in$ $L_{{i-1}}$ not TRANSIT}{\label{alg1-l13}
      $H_{c}$ $\gets$ current state of $\mathcal{A}(\alpha)$

            \If{there is no $\mathcal{A}(\alpha)' \in L_{i}$ whose current state is $H_c$}{
                 add a copy of $\mathcal{A}(\alpha)$ to $L_{i}$\label{alg1-l16}
            }
        }

                increase $P(L_{i})$ by $P(L_{i-1})$ $\cdot$  $P(E'_i | \mathcal{M}_{\mathcal{I}_{\alpha}}, t_i)$, update $\mathcal{L}_{i}$ \label{alg-l18}

            }
        }
    }
\Return $\mathbb{E}_{\hat{\bm{S}}\sim{P(\hat{\bm{S}} | \mathcal{M}_{\mathcal{I}_{\alpha}}, n)}}(\mathrm{sp}(\alpha | \hat{\bm{S}}))$

\end{algorithm}
\vspace{-3mm}

Then we sequentially consider every pair of possible active finite automaton list $L_{i-1}$ and possible event set $E'_{i}$ produced by the generative model $\mathcal{M}_{\mathcal{I}_{\alpha}}$ for every time stamp $t_i$ where $1$${\leqslant}$ $i$ ${\leqslant}$ ${\mathrm{len}{(\bm{S})}}$~(Line~\ref{alg-l4}--\ref{alg-l7}).
For each active $\mathcal{A}(\alpha)$ in $L_{i-1}$, we invoke the state transitive function by consuming its current state $H_c$ and the event set $E_{i}'$ and acquire the output state $H_t$~(Line~\ref{alg-l8}--\ref{alg-l10}). If a transition occurs, we check both $H_c$ and $H_t$ for automaton management. In particular, if $H_t$ reaches the sink state $H_{k}^{\alpha}$, we obtain a minimal occurrence of $\alpha$ and increase the expected support of $\alpha$ by $P(L_{i-1})$ $\cdot$  $P(E'_i | \mathcal{M}_{\mathcal{I}_{\alpha}}, t_i)$~(Line~\ref{alg-l11}--\ref{alg-l13}).
Otherwise, we add the updated $\mathcal{A}(\alpha)$ to $L_{i}$. We meanwhile check whether $H_c$ is a source state and add an initialized automaton to $L_{i}$ if it is~(Line~\ref{alg-l16}--\ref{alg-l17}). For the rest automatons without state transitions, we will determine whether they can be added to $L_{i}$ since we are interested in minimal occurrence of episode~(Line~\ref{alg1-l13}--\ref{alg1-l16}).
Finally we update the probability of the appearance of active automatons $L_{i}$ for further computations~(Line~\ref{alg-l18}).

For example, consider the episode \ev{abc} and the generative model $\mathcal{M}_{\ev{ab}}$ as shown in Fig.~\ref{fig:generative}.
When $i$ $=$ $3$, the only possible active automaton list $L_{2}$ $\in$ $\mathcal{L}_{2}$, having three active automatons, contains $\{H_{0}^{\ev{abc}}$, $H_{1}^{\ev{abc}}$, $H_{2}^{\ev{abc}}\}$. Here $H_{i}^{\ev{abc}}$ denotes the current state of the corresponding automaton. For possible event sets, there are two possible event sets, i.e. $E'_{3}$ $=$ $\{\ev{b}\}$ and $E''_{3}$ $=$ $\{\ev{b}, \ev{c}\}$, with the probability of 0.5 of each, respectively.
Considering every pair of the active automaton and the event set, we detect $T(H_{2}^{\ev{abc}}, E''_{3})$ derives the SINK state. We thus add the expected support of the episode $\ev{abc}$ under the generative model $\mathcal{M}_{\ev{ab}}$ by $P(L_2)$$\cdot$$P(E''_{3} | \mathcal{M}_{\ev{ab}}, t_{3})$ $=$ $0.5$. Then we update the probability of active automatons and derive  $\mathcal{L}_{3}$$=$$\{L_{3}\}$$=$$\{H_{0}^{\ev{abc}}, H_{2}^{\ev{abc}}\}$ with $P(L_3)=1$.

\subsection{Time Complexity Analysis}
Here we analyze the time complexity of the algorithm, i.e., Algorithm~\ref{alg:expect}, for expected support calculation given a specific generative model. We only discuss the worst case for simplicity.
Given a specific generative model $\mathcal{M}_{\mathcal{I}_{\alpha}}$ of an episode $\alpha$, for each time stamp $t_i$, there are at most
$|\alpha|$ active automatons in each active automatons list $L_i$, $2^{|\overline{\mathcal{I}_\alpha}|}$ possible event set $E_j¡¯$ and $2^{|\alpha|}$ possible lists of active automatons, respectively. Hence the time complexity for precessing a time stamp $t_i$ in the wort case is $O(|\alpha|\cdot2^{|\overline{\mathcal{I}_\alpha}|+|\alpha|})$. Since there are all together $n$ time stamps on the event sequence $\bm{S}$, the overall time complexity of Algorithm~\ref{alg:expect} will be $O(n\cdot|\alpha|\cdot2^{|\overline{\mathcal{I}_\alpha}|+|\alpha|})$ where $n$ denotes the length of the event sequence. Though it is exponential in $|\overline{\mathcal{I}_\alpha}|$ and $|\alpha|$, this number might not be very large as usually the episode to be checked has limited length and so is $|\overline{\mathcal{I}_\alpha}|$.

\section{Experiments} \label{sec:experiments}\vspace{-2mm}
In this section, we present the results of our experimental studies.

\vspace{-3mm}
\subsection{Datasets and Experiment Settings} \label{dataset}
We conducted the experiments on both synthetic and real-world datasets, and Table~\ref{tab:datasetcha} illustrates their statistics.

The \textbf{SYN} dataset refers to a synthetic dataset which is generated as follows.
We use an alphabet of $52$ events, i.e., $\ev{a} \cdots \ev{z}$ and $\ev{A} \cdots \ev{Z}$ for generating a sequence whose length is set to $10,000$.
In such sequence, we planted two episodes and a high frequency noise event. The first episode \ev{abc} with no gaps was embedded $300$ times into randomly selected time stamps. The second episode, a $4$-episode $\ev{defg}$, can have a gap between any two consecutive events. The gap is a positive integer drawn from a truncated normal distribution $\mathrm{N}(2, 2)$. The episode $\ev{defg}$ was randomly planted $300$ times in the sequence as well. These two patterns might overlap but there are no generative rules between them. Third, we planted a high frequency event \ev{X} with $p_{\mathit{ind}}(\ev{X})$ $=$ $0.3$ into randomly selected time stamps of the sequence. Such event might become a free-rider event that doped with the embedded patterns. Finally, we filled the sequence using the rest events which were generated from a uniform distribution. They are considered as random noises. As a consequence, all the episodes rather than the non-empty subepisodes of $\ev{abc}$ and $\ev{defg}$ are considered as free-riders in such dataset.

\begin{table}\scriptsize
\centering\vspace{-8mm}
\caption{Statistical information of datasets} \label{tab:datasetcha}
\begin{tabular}{ccccccc}
\toprule
 \multirow{2}{*}{\textbf{Dataset}} & {\textbf{Sequence}} & {\textbf{Events in}} & {\textbf{Avg. events}}  & {\textbf{Avg. occurrences}}
  \\ \multirow{2}{*}
  {}&{\textbf{Len.}}&{\textbf{alphabet}}&{\textbf{per timestamp}}&{\textbf{per event}}\\
  \midrule
  SYN & $10,000$ & $52$ & $1.4$ & $267.3$ \\
  STK & $9,334$ & $50$& $5.55$ & $1,037$ \\
  JMLR& $75,645$ & $3,846$ & $1.0$ &$19.67$ \\
 \bottomrule
\end{tabular}
\end{table}
\vspace{-5mm}

The \textbf{STK} dataset consists of daily prices of stocks from Chinese stock market. 
It includes $50$ blue chip stocks with their price information over 26 years from December 19th, 1990 to July 14th, 2016.
The events for a stock are generated by the increase ratio of price on each trading day. If the ratio is positive, we generate an increase event for such stock on the corresponding day. For example, if the stock of Bank of China increases on December 19th, 1990, we will generate an event as ``Bank of China+'' for that day. Otherwise, we will not produce event for the stock of Bank of China on that day. Similar operations were performed for all the 50 stocks in such dataset.

The \textbf{JMLR} dataset consists of the abstracts of the papers published on the Journal of Machine Learning Research which was adopted in~\cite{tatti2015ranking}. We treat each word as an event and built an event sequence by connecting every sentence together. Such dataset was preprocessed by stemming the words and removing the stop words. The details can be found in~\cite{tatti2015ranking}.

We compare our EDP model against the following state-of-the-art methods:

 1. \textbf{PRT}~\cite{tatti2015ranking}: A partition model that divides an episode into two consecutive sub-episodes and detects whether its frequency can be explained by the two sub-episodes.

 2. \textbf{SkOPUS}~\cite{petitjean2016skopus}: A partition model that compares the frequency of a pattern with its re-ordering candidates consisting of two sub-pattern partitions.

 3. \textbf{EGH}~\cite{laxman2005discovering}: An independence baseline that connects the significance of an episode with a stationary hidden markov model.

 4. \textbf{IND}: The degraded version of the proposed EDP model which considers every event is a random event.

Since all the compared methods require a set of candidate patterns, we start from a set of frequent episodes, denoted as $\mathcal{F}$, in which we require $|\Omega_{\alpha}|$ $>$ $1$ for every $\alpha$ $\in$ $\mathcal{F}$.\footnote{Here we do not take any frequent event or the episode consisting of single event into account.}
PRT and SkOPUS are designed for multiple sequences rather than a single event sequence, we thus transfer our event sequence into database of short sequences by segmenting it with fixed length sliding windows to fit the methods. For mining frequent episodes and deriving $\mathcal{F}$, we adopted the DFS algorithm~\cite{fem:achar2013pattern}, which is a state-of-the-art minimal occurrence based frequent episode mining algorithm.

\subsection{Results on data with known patterns}

We first demonstrate our results on the SYN dataset having known patterns. In particular, we mined $\mathcal{F}$ by setting $min\_sup = 200$ and $\delta$ $=$ $12$ and finally obtained $|\mathcal{F}|$ $=$ $325$. Such setting ensures the embedded real patterns are contained in $\mathcal{F}$. Our purpose is to check whether the compared methods can unearth these planted patterns and rank them as high as possible.\\

\makeatletter\def\@captype{table}\makeatother
\begin{minipage}{.45\textwidth}\scriptsize
\caption{Precision@$\mathit{k}$ on SYN dataset.}\vspace{-3mm}\label{tab:compare}
\begin{tabular}{c|rrrrr}
    \toprule
    \multirow{2}[4]{*}{\textbf{Top k}} & \multicolumn{5}{c}{\textbf{Precision@k}} \\
\cmidrule{2-6}          & \multicolumn{1}{c}{\textbf{IND}} & \multicolumn{1}{c}{\textbf{EGH}} & \multicolumn{1}{c}{\textbf{SkOPUS}} & \multicolumn{1}{c}{\textbf{PRT}} & \multicolumn{1}{c}{\textbf{EDP}} \\
    \midrule
    1     & 100.0\% & 0.0\% & 100.0\% & 100.0\% & 100.0\% \\
    2     & 100.0\% & 50.0\% & 100.0\% & 100.0\% & 100.0\% \\
    3     & 100.0\% & 66.7\% & 100.0\% & 100.0\% & 100.0\% \\
    4     & 100.0\% & 75.0\% & 100.0\% & 100.0\% & 100.0\% \\
    5     & 80.0\% & 60.0\% & 100.0\% & 100.0\% & 100.0\% \\
    6     & 66.7\% & 50.0\% & 100.0\% & 83.3\% & 100.0\% \\
    7     & 71.4\% & 57.1\% & 85.7\% & 71.4\% & 100.0\% \\
    8     & 75.0\% & 62.5\% & 75.0\% & 62.5\% & 100.0\% \\
    9     & 66.7\% & 66.7\% & 66.7\% & 66.7\% & 100.0\% \\
    10    & 60.0\% & 60.0\% & 60.0\% & 60.0\% & 100.0\% \\
    11    & 54.5\% & 54.5\% & 54.5\% & 63.6\% & 100.0\% \\
    12    & 50.0\% & 58.3\% & 50.0\% & 66.7\% & 100.0\% \\
    13    & 46.2\% & 53.8\% & 46.2\% & 69.2\% & 100.0\% \\
    14    & 42.9\% & 57.1\% & 50.0\% & 71.4\% & 100.0\% \\
    15    & 40.0\% & 60.0\% & 53.3\% & 73.3\% & 100.0\% \\
    \bottomrule
    \end{tabular}%
\end{minipage}
\quad
\makeatletter\def\@captype{table}\makeatother
\begin{minipage}{.45\textwidth}\scriptsize

\caption{Details of top $15$ episodes.}\vspace{-3mm}\label{tab:syndetail}
\begin{tabular}{c|rrrr}
    \toprule
    \textbf{Top $k$} & \multicolumn{1}{c}{\textbf{SkOPUS}} & \multicolumn{1}{c}{\textbf{PRT}} & \multicolumn{1}{c}{\textbf{EDP}} \\
    \midrule
    1      & \ev{d$\rightarrow$e$\rightarrow$f$\rightarrow$g} & \ev{d$\rightarrow$e$\rightarrow$f$\rightarrow$g} & \ev{e$\rightarrow$f$\rightarrow$g} \\
    2    & \ev{a$\rightarrow$b$\rightarrow$c} & \ev{d$\rightarrow$e$\rightarrow$g} & \ev{f$\rightarrow$g} \\
    3      &\ev{d$\rightarrow$e$\rightarrow$f} & \ev{d$\rightarrow$f$\rightarrow$g} & \ev{e$\rightarrow$f} \\
    4      &\ev{e$\rightarrow$f$\rightarrow$g} & \ev{e$\rightarrow$f$\rightarrow$g} & \ev{e$\rightarrow$g} \\
    5      & \ev{d$\rightarrow$e$\rightarrow$g} & \ev{d$\rightarrow$e$\rightarrow$f} & \ev{d$\rightarrow$e$\rightarrow$f} \\
    6      & \ev{\bluefont{c$\rightarrow$X$\rightarrow$d}} & \ev{\bluefont{a$\rightarrow$c$\rightarrow$d}} & \ev{d$\rightarrow$f} \\
    7      & \ev{\bluefont{b$\rightarrow$c$\rightarrow$X$\rightarrow$d}} & \ev{\bluefont{a$\rightarrow$b$\rightarrow$c$\rightarrow$d}} & \ev{d$\rightarrow$e} \\
    8      & \ev{\bluefont{b$\rightarrow$c$\rightarrow$d}} & \ev{\bluefont{b$\rightarrow$c$\rightarrow$d}} & \ev{a$\rightarrow$b}\\
    9      & \ev{\bluefont{b$\rightarrow$X$\rightarrow$d}} & \ev{a$\rightarrow$b$\rightarrow$c} & \ev{b$\rightarrow$c} \\
    10     & \ev{\bluefont{a$\rightarrow$b$\rightarrow$c$\rightarrow$d}} & \ev{\bluefont{a$\rightarrow$b$\rightarrow$d}} & \ev{a$\rightarrow$c} \\
    11     & \ev{\bluefont{a$\rightarrow$b$\rightarrow$X$\rightarrow$d}}& \ev{f$\rightarrow$g} & \ev{a$\rightarrow$b$\rightarrow$c} \\
    12    & \ev{\bluefont{a$\rightarrow$c$\rightarrow$X$\rightarrow$d}} & \ev{e$\rightarrow$f} & \ev{d$\rightarrow$e$\rightarrow$f$\rightarrow$g} \\
    13     & \ev{f$\rightarrow$g} & \ev{e$\rightarrow$g} & \ev{d$\rightarrow$f$\rightarrow$g} \\
    14     & \ev{e$\rightarrow$f} & \ev{d$\rightarrow$e} & \ev{d$\rightarrow$e$\rightarrow$g} \\
    15    & \ev{e$\rightarrow$g} & \ev{d$\rightarrow$f} & \ev{d$\rightarrow$g} \\
    \bottomrule
    \end{tabular}%
\end{minipage}\\

We explored every compared method on the $\mathcal{F}$ and ranked the episodes by the deviation between the observed support and the expectation. For our EDP and IND model, we set $\mathit{min}\_\mathit{lift}$ $=$ $1$ and ranked the episodes whose Lift exceeds $\mathit{min}\_\mathit{lift}$ with descending order.
For other methods we directly ranked by the deviation used in their papers. We report precision at $k$ in Table~\ref{tab:compare}, that is, proportions of embedded patterns that are included in the top-$k$ patterns returned by each approach. Remember that non-empty subepisodes of $\ev{abc}$ and $\ev{defg}$ except the single events are real patterns based on our definition as there exists underlying generative rules to explain it.
As a result, there could be $15$ embedded patterns in the dataset. What we concern about is the ability of each approach to recognize these embedded patterns.

From the table, we observe the independence models, i.e. EGH and IND, do not perform well as expected, while the partition based models are more effective. Our EDP model, with the highest precision, ranks all the embedded patterns at top 15 rankings and significantly outperforms other methods.
PRT and SkOUPS begin to report false positive patterns after 5th and 6th, respectively.
In addition, we observe EDP provides 20 outputs whose Lift is greater than the threshold which was 1 in our setting. However we observe a large gap between the lift measure of the 15th and 16th in the results of EDP, which doubles the gap between the 1st and 15th. Though 16th-20th are false positives, we argue that we can filter the additional noises by finer tuning the threshold based on the gaps in the lift and our method already returns the targets on top 15.

Next we look closer to the results returned by the three partition models. Table~\ref{tab:syndetail} demonstrates the top 15 episodes produced by the three partition based models. Among them, we mark the false positive patterns in blue font. From the table we observe that SkOPUS cannot handle the high frequency noise event, i.e. \ev{X}, very well, and \ev{X} could be doped inside patterns. On the other hand, SkOPUS may underestimate the expectation of patterns with longer length as their re-ordering candidates may be rare in the original sequence. Thus, the top 5 of SkOPUS are 3-episode or longer, and it can return \ev{defg} and \ev{abc} very early.
The ineffectiveness of PRT comes from its incorrectness in identifying the event from other patterns. For example, it fails to recognize \ev{d} is a free-rider event to the pattern \ev{abc}. However, our EDP can correctly filter such episodes.
As the fact that EDP can recognize more complicated redundant patterns, we conjecture the reason is that our partition model takes non-prefix episodes into account during the partition.

\begin{figure} [htbp]
\centering\vspace{-5mm}
\includegraphics[height=3cm, width=6.5cm]{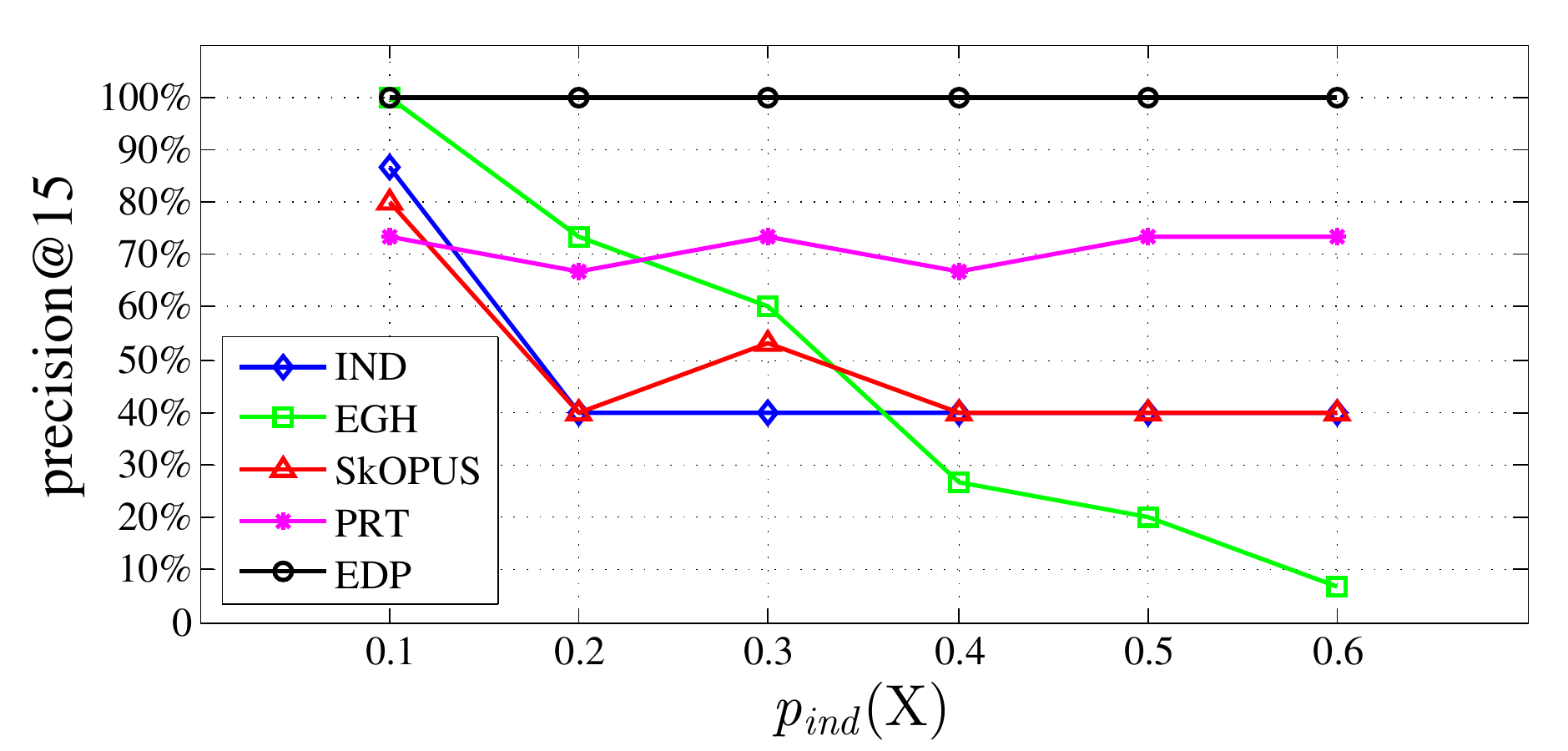}
\vspace{-3mm}
\caption{The effect of $p_{\mathit{ind}}($\ev{X}$)$.}\label{fig:sensitivity}
\vspace{-5mm}
\end{figure}

We next tune the probability of the embedded high frequency noise event, i.e. $p_{\mathit{ind}}(\ev{X})$, from $0.1$ to $0.6$ and visualize the Precision@$15$ of all compared methods as shown in Fig.\ref{fig:sensitivity}. From the figure, we observe that the probability of \ev{X} has few effects on EDP. It always achieves the highest precisions with $100\%$. Second, the measure of IND, SkOPUS and PRT may vary around $40\%$ to $75\%$ as $p_{\mathit{ind}}(\ev{X})$ varies. The only special case happens when $p_{\mathit{ind}}(\ev{X})$ $=$ $0.1$, both IND and SkOPUS achieve more than $80\%$ in precision. EGH, on the other hand, is much more sensitive to the $p_{\mathit{ind}}(\ev{X})$, and its performance drops significantly as the $p_{\mathit{ind}}(\ev{X})$ increases. However, it performs well when the appearance probability of \ev{X} is low.

As the synthetic experiments clearly show that IND and EGH overall are less effective than the others, we do not include them in the following experiments.

\subsection{Results on real-world dataset}

\textbf{Results on STK dataset.} We start from the most frequent $1,000$ episodes mined from the STK dataset by setting $\mathit{min}\_\mathit{sup}$ $=$ $200$ and $\delta$ $=$ $5$. 
We also ranked each episode by the deviation between the observed support and the expectation, and meanwhile we set $\mathit{min}\_\mathit{lift}$ to 1 to screen redundant episodes for our method. We were surprised to find that each approach returned very different results on such dataset. Since the episodes related to stocks are not easy to understand, we design the following measure to demonstrate the results.

\begin{table}[!htbp]\scriptsize
  \centering
  \caption{The percentage of the most frequent 10 events in the top $k$ episodes in STK.} \label{tab:stkcompare}
    \begin{tabular}{c|rrr}
    \toprule
    \textbf{Top $k$} & \multicolumn{1}{c}{\textbf{SkOPUS}} & \multicolumn{1}{c}{\textbf{PRT}} & \multicolumn{1}{c}{\textbf{EDP}} \\
    \midrule
    1     & 33.3\% & 100.0\% & 0.0\% \\
    2     & 50.0\% & 100.0\% & 0.0\% \\
    3     & 44.4\% & 100.0\% & 0.0\% \\
    4     & 33.3\% & 100.0\% & 0.0\% \\
    5     & 33.3\% & 100.0\% & 20.0\% \\
    6     & 33.3\% & 100.0\% & 33.3\% \\
    7     & 33.3\% & 100.0\% & 28.6\% \\
    8     & 33.3\% & 100.0\% & 37.5\% \\
    9     & 33.3\% & 100.0\% & 33.3\% \\
    10    & 33.3\% & 100.0\% & 30.0\% \\
    \bottomrule
    \end{tabular}%
  \label{tab:addlabel}\vspace{-5mm}
\end{table}

For the top $k$ episodes given by each approach, we exhibit the percentage of the top $10$ frequent events are included in. The reason we choose top $10$ frequent events is that they have relatively high occurrence probabilities ranging from $0.2$ to $0.26$, which is twice to the average occurrence probability of events in such dataset~(see Table~\ref{tab:datasetcha}). 
From the results on the synthetic dataset, we have known that real patterns may dope with some highly frequent but not related events, which makes the episode become a free-rider. As a result, we argue that such ratio is higher may indicate that there might be more possibility to have free-rider episodes in the top $k$ episodes.

Table~\ref{tab:stkcompare} demonstrates the results, and we observe PRT achieves surprisingly high percentages. EDP outperforms SkOPUS especially when $k$ is less than 5.
To validate further, we check the top episodes produced by every method.
We find that, as an example, the highest ranking episode returned by PRT indicates that the increase of a software company~(SH.600100+) is followed by the increase of a security corporation~(SH.600109+) and then leads to the increase of a metallurgical enterprise~(SH.600111+). All the three events belong to different industry sectors and are included in top ten frequent events. In fact, from our common sense, we can hardly believe these three stocks have convincing correlations with each other. The results of SkOPUS is similar with that in the SYN dataset that the underlying patterns are always doped with the most frequent event, while the results provided by EDP are clearly different. For example, the episode ``\textbf{Industrial and Commrcl Bank of China+}''$\rightarrow$``\textbf{Bank of China+}'' and ``\textbf{Minsheng Bank+}''$\rightarrow$``\textbf{China Merchants Bank+}'' rank at $1^{\mathrm{st}}$ and $2^{\mathrm{rd}}$, respectively.
Such patterns are more convincing and easy to understand from the stock names. The stocks in the same episode may have implicit correlations as they come from the same industry sector. We also find more examples from the results returned by EDP model but further discussions on possible related stocks are out of the scope of this paper.

\begin{table*}\scriptsize
\centering\vspace{-10mm}
\caption{Top 10 output episodes on JMLR dataset.} \label{tab:text}
\begin{tabular}{cccccccccc}
\toprule
  \textbf{Top k} &  \textbf{SkOPUS} & \textbf{PRT} & \textbf{EDP} \\
  \midrule
  {1} &  {support$\rightarrow$vector$\rightarrow$machin~(138)} & {support$\rightarrow$vector~(168)} & {support$\rightarrow$vector~(168)}  \\
  {2} & {support$\rightarrow$vector~(168)} & {support$\rightarrow$vector$\rightarrow$machin~(138)} & {real$\rightarrow$world~(78)}  \\
  {3} & {vector$\rightarrow$machin~(151)}& {support$\rightarrow$machin~(142)}  & {support$\rightarrow$vector$\rightarrow$machin~(138)}  \\
  {4} & {support$\rightarrow$machin~(142)} & {vector$\rightarrow$machin~(151)} & {support$\rightarrow$machin~(142)}  \\
  {5} & {data$\rightarrow$set$\rightarrow$model~(79)} & {real$\rightarrow$world~(78)} &{vector$\rightarrow$machin~(151)}  \\
  {6} & {base$\rightarrow$algorithm$\rightarrow$base~(78)} & {featur$\rightarrow$select~(101)} & {bayesian$\rightarrow$network~(95)}  \\
  {7} & {data$\rightarrow$set$\rightarrow$data$\rightarrow$set~(76)} & {bayesian$\rightarrow$network~(95)} & {solv$\rightarrow$problem~(84)}  \\
  {8} & {paper$\rightarrow$learn$\rightarrow$algorithm~(75)} & {problem$\rightarrow$solv~(79)} & {data$\rightarrow$set~(298)$^{\star}$}  \\
  {9} & {set$\rightarrow$learn$\rightarrow$set~(75)} & {bayesian$\rightarrow$model~(92)} & {machin$\rightarrow$learn~(77)$^{\star}$}  \\
  {10} & {real$\rightarrow$data~(97)} & {solv$\rightarrow$problem~(84)} & {real$\rightarrow$data~(97)}  \\
 \bottomrule
\end{tabular}
\end{table*}\vspace{-5mm}

\textbf{Results on JMLR dataset.}
For the JMLR dataset, we are given $1,023$ frequent episodes in $\mathcal{F}$ by setting $\mathit{min}\_\mathit{sup}$ $=$ $50$ and $\delta=12$. Then we explored the three partition models on such dataset. We visualize the top $10$ episodes ranked by each method in Table~\ref{tab:text}. In such table, the number in the bracket is the support of corresponding episode. Since there are no ground truth patterns in such data, we can hardly compare which method is better in a quantitative manner. However, we find the following interesting observations.
First, confirm the ones in the previous comparisons, SkOPUS outputs free-rider episodes like ``\textbf{data}$\rightarrow$\textbf{set}$\rightarrow$\textbf{model}''
and ``\textbf{base}$\rightarrow$\textbf{algorithm}$\rightarrow$\textbf{base}'' at very early. In addition, SkOPUS seems to have negative correlations with support since it prefers longer patterns. We know the JMLR dataset is a sequence without simultaneous events, hence support based on minimal occurrence holds anti-monotonicity.
PRT performs well on such data. Every pattern demonstrated in the table is meaningful, and all of them have lower frequency which is less than $200$.
Our EDP also provides meaningful results and has few correlations with pattern support. It can rank both high frequency as well as low frequency episode high, e.g. ``\textbf{data}''$\rightarrow$``\textbf{set}''~(298) and ``\textbf{machin}''$\rightarrow$``\textbf{learn}''~(77). We mark them by asterisk in the table.
Since the basic idea of EDP is checking whether the observed support can be simulated by the random sequences generated by specific partitions. Episodes with high or low support may have possibility to pass such test. We conjecture it may be the reason why EDP can return both general and specific patterns.

\vspace{-3mm}
\subsection{Scalability}

\begin{figure} [htbp]
\centering\vspace{-7mm}
\includegraphics[height=4cm, width=6cm]{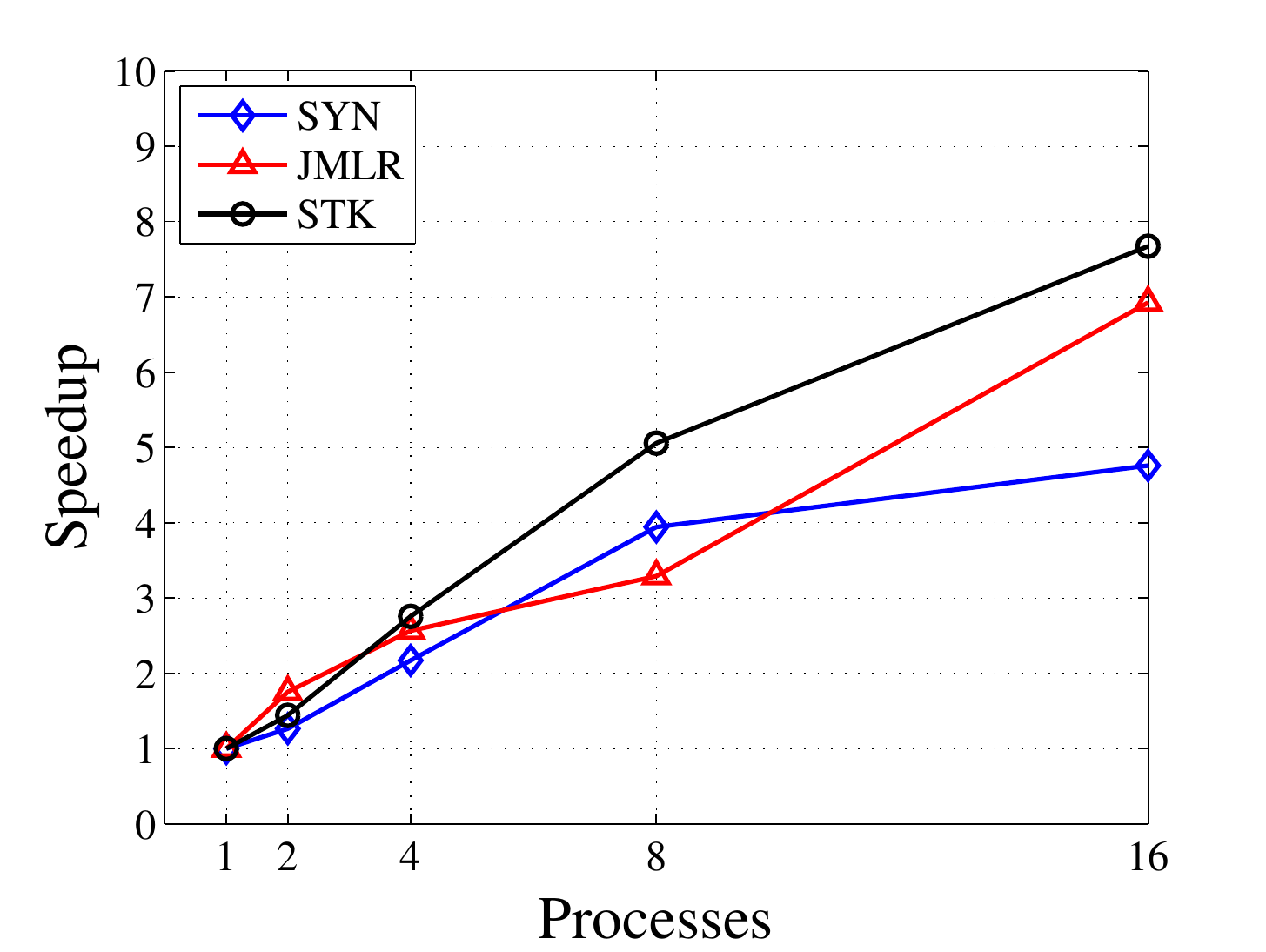}
\vspace{-3mm}
\caption{The scalability of EDP.}\label{fig:scalability}
\vspace{-6mm}
\end{figure}

Finally we evaluate the efficiency of EDP. Since EDP can be highly distributed, we mainly focus on the scalability of EDP by varying the number of processes. To this end, we implemented EDP with Python 3.5, distributed each partition model of an episode into different processes and verified its scalability on a shared memory computing system. The system equipped with two Intel Xeon E5-2620 CPUs and 128GB RAM running on Linux CentOS 6.5. Fig.~\ref{fig:scalability} demonstrates the results. From the figure we can see EDP can achieve significant speedup as the number of processes increases on all datasets we investigated. Among them, EDP holds the most advanced speedup on the STK dataset. It is because we adopt small $\delta$ when performing mining such that the episodes on STK have less length and smaller alphabet.

\vspace{-5mm}\section{Conclusion}\label{sec:conclusion}\vspace{-3mm}
In this paper, we proposed EDP model for reducing redundancy of frequent episodes in long single sequence. EDP first separates an episode into informative and random events, then builds a generative model for random sequences based on the partitions. Next we define the expected support of an episode if it is considered as a free-rider with these random sequences and devise an efficient algorithm to compute.
Finally, the redundancy of an episode is determined by the deviation between the observation and the estimated expectation. Experimental results on synthetic and real-world datasets clearly exhibit the effectiveness of the proposed method. \\

\small{\textbf{Acknowledgements.} \ \  This research is supported by National key R\&D program of China (No. 2017YFB1002104), National Natural Science Foundation of China (No.61602438, 91546122, 61573335), Guangdong provincial science and
technology plan projects (No. 2015B010109005).}\vspace{-3mm}

\bibliographystyle{plain}
\bibliography{edp}

%
%
%
%
%
%
%

\end{document}